# Effects of Voice-Based Synthetic Assistant on Performance of Emergency Care Provider in Training

Praveen Damacharla [a], Parashar Dhakal [a], Sebastian Stumbo [b], Ahmad Y. Javaid [a, *],
Subhashini Ganapathy [b], David A. Malek [c], Douglas C. Hodge [d, a] and Vijay Devabhaktuni [a]

[a] Department of EECS, The University of Toledo, Toledo, Ohio, USA
[b] Department of Biomedical, Industrial & Human Factors, Wright State University, Dayton, Ohio, USA
[c] Wright State Research Institute, Wright State University, Dayton, Ohio, USA
[d] National Center for Medical Readiness, Wright State University, Dayton, Ohio, USA

**Abstract:**

As part of a perennial project, our team is actively engaged in developing new synthetic assistant (SA) technologies to assist in training combat medics and medical first responders. It is critical that medical first responders be well trained to deal with emergencies more effectively. This would require real-time monitoring and feedback for each trainee. Therefore, we introduced a voice-based SA to augment the training process of medical first responders and enhance their performance in the field. The potential benefits of SAs include a reduction in training costs and enhanced monitoring mechanisms. Despite the increased usage of voice-based personal assistants (PAs) in day-to-day life, the associated effects are commonly neglected for a study of human factors. Therefore, this paper focuses on performance analysis of the developed voice-based SA in emergency care provider training for a selected emergency treatment scenario. The research discussed in this paper follows design science in developing proposed technology; at length, we discussed architecture and development and presented working results of voice-based SA. The empirical testing was conducted on two groups as user study using statistical analysis tools, one trained with conventional methods and the other with the help of SA. The statistical results demonstrated the amplification in training efficacy and performance of medical responders powered by SA. Furthermore, the paper also discusses the accuracy and time of task execution (t) and concludes with the guidelines for resolving the identified problems.

---

* Corresponding author e-mail address: ahmad.javaid@utoledo.edu

1. **Introduction**

The emergency care provider (ECP) provides treatment at the point of injury and contributes significantly to the success of a patient's survival in case of emergencies in both civilian and military domains. Recent studies point out that the number of lost lives due to medical errors is at 251,000 per year, and yet the estimation is considered to be the lowest number. To this end, medical errors are the third-largest cause of death in the United States, even though they are preventable with increased efforts (Makary & Daniel, 2016). It is also reported that most common medical errors occur during emergency treatment or transfer of care between different care providers (Hobgood, Hevia, Tamayo-Sarver, Weiner, & Riviello, 2005; Moorman, 2007). Per our estimations, the fatalities caused by medical errors in emergency response account for 50% of the total fatalities. Therefore, to minimize these errors, a carefully monitored training system for ECPs (for example, live, virtual, constructive environment) is envisaged to efficiently deal with medical emergencies (Henninger et al., 2008).

*1.1 Research Background*

In healthcare, most of the errors take place during information exchange between patients and caretakers due to procedural execution mistakes, inadequate documentation, and poor communication (Nagpal et al., 2012). With the application of SA in the medic training phase, it is possible to reduce the number of these errors. Here, the role of the SA is to monitor the training process in real time and correct the trainee whenever a mistake is committed. The enhanced training module also improves the communication and procedural tactics of medics, thereby reducing the size of the error window (M. He, Wang, Todd, Zhao, & Kezys, 2007). Therefore, an SA built with custom features and voice-triggered communication significantly augments a medic's capabilities to accomplish a task. Moreover, a voice-based interaction is hypothesized to be better than other modes of communication in making the treatment process quick, efficient, and user-friendly (Jiang et al., 2015; Moorthy & Vu, 2014). Some of the other technologies, such as a pocket card with a barcode reader, mobile applications, and medical software with touch screens, are also used for information transfer between different medical teams. However, given their inherent drawbacks of being time-consuming and distracting users from primary goals, these interfacing technologies are ineffective and constrained in assisting in complex tasks (J. He et al., 2014). Therefore, a real-time, voice-based SA would most likely address all the above-stated drawbacks. A voice query additionally leverages a sophisticated natural language processing (NLP) Q&A system to produce a natural language response to the user (Hauswald et al., 2015). One of the major problems with the use of speech communication has been substantial recognition errors. With the advent of new cloud-based NLP systems, this problem has been solved to a great extent.

Existing studies proved that the replacement of humans with intelligent synthetic agents to train aviation teams would not attribute any loss in training efficiency (Demir et al., 2015; Zachary, Santarelli, Lyons, Bergondy, & Johnston, 2001). In a generic sense, a synthetic agent or intelligent software agent is defined as "Complex software entities that carry out some set of operations on behalf of a user with some degree of autonomy" (Franklin & Graesser, 1997). SA is a multi-layered intelligent system that sits on top of other services or applications and performs actions to fulfill the user's intent. Here, intent could be a user command to perform an action. SA makes use of a core set of technologies, such as machine learning, speech recognition, dialogue management, language generation, text-to-speech synthesis, data mining, analysis, inference, and personalization for performing tasks (Sarikaya, 2017; Yorke-Smith, Saadati, Myers, & Morley, 2009). SAs can be useful to a user by providing easy access to personal/external data, web services, and other applications, such as finding documents, locating places, and playing music. In addition, they can also provide the user with notifications and alerts, such as alarms, meeting reminders, and meeting locations. These features increase user productivity by assisting them in work and life management (Lee, Kim, & Lee, 2015; Mitchell, Caruana, Freitag, McDermott, & Zabowski, 1994; Myers et al., 2007; Zhang et al., 2016). The evolution of a typical SA started in 1996 and gradually permeated into a software simulation environment (Bunch et al., 2004). With the advancement in NLP and image processing algorithms, SA got the upgrade it needed, and this favored easy interfacing with the surrounding environment. SA research further synergized with advanced cognitive concepts to handle high-degree complex tasks (Ball et al., 2010). Owing to continuous efforts in cementing perfection within the NLP techniques, the industry has created several interactive voice-based personal assistants (PAs). Apple Siri, Google Assistant, Amazon Alexa, Microsoft Cortana, and IBM Watson are some of the commercially available intelligent PAs that provide proactive and reactive assistance to users (Sun et al., 2016). Nowadays, PAs have become an essential capability in most smartphones, tablets, laptops, desktops, and hands-free devices. In general, the user interacts with PAs through natural language commands, and the PA performs an action based on the user's intent (Sarikaya, 2017). These outcomes motivated us to design an interactive voice interface SA for ECPs.

*1.2 Research hypothesis*

We hypothesize that a voice-based SA can play a crucial role in the rapid training of ECPs, which can improve their performance and efficiency in the field. To achieve this, we developed an interactive, autonomous voice-based SA to train ECPs. An SA corrects and recommends steps to resolve mistakes committed by a trainee in medical treatment and further enables a closely monitored training system to improve ECPs efficiency. A human trainer can observe only one trainee at a time. On the other hand, SAs can monitor multiple trainees simultaneously, deeming SAs to be a better alternative. We hypothesize that the application of SAs renders a higher degree of monitoring of each trainee at every phase of training. This helps the human participant to correct the errors during the training phase and has a higher chance of minimizing human errors during execution. The paper is subsequently divided into the following sections: voice-based synthetic assistant architecture and implementation, the methodology of research for user evaluation, results and discussion, and conclusion.

2. **Voice-Based Synthetic Assistant**

A voice-based SA (VBSA) uses voice interaction as the primary communication mode. Many SA techniques that are task-specific are being implemented by different organizations (Ball et al., 2010; Zachary et al., 2001). Indeed, the developed VBSA is also a task-specific SA. Subject matter expert, Colonel Douglas C. Hodge, used his experience and published work to define the specifications for an ECP training-related task. This task was then used to develop the required SA architecture (Moorman, 2007; Ong, BiomedE, & Coiera, 2011; Stephen Harden; Memphis, 2011). Concerning subject matter expert specifications, we have identified the following requirements:

a) VBSA task "assist ECP with error detection and recommend steps to resolve errors."

b) Human-related task "provide emergency care to the patient and simultaneously report to SA."

c) Interface mode "do not allow any hindrances to ECP's motor skills while treating the patient."

To meet these specifications, the paper classifies (1) process flow monitoring, processed data storage, and training progress statistics under the VBSA task, and (2) patient care. Communication with the SA occurs under the human task. The voice interface has become a potential mode of the interface without compromising human gross motor skills in medical treatment. The VBSA is elaborated in two parts, with a description of design and architecture of VBSA in part I, followed by the realization of architecture with conventional technologies and operation flow in part II. Although the scenario selected for testing the developed VBSA is narrow (only the massive hemorrhage scenario), the VBSA is developed for the complete Massive hemorrhage, Airway blockage, Respiration, Circulation, and Hypothermia (MARCH) protocol, also known as MARCHp. This protocol is used in most combat medic training programs. The applications of VBSA can be used in many parts of medic training that include training at the point of injury, training for transfer of care, and in-hospital care. It is essential to understand that the list of potential applications is relatively extensive. This technology can be expanded to several other application areas ranging from training in medicine (e.g., monitoring pre-surgery checklist) to simple training in education or service industry (e.g., assembling parts in an automobile).

*2.1 Architecture*

The developed SA monitors the training progress of an ECP in real time and helps the ECP resolve errors committed, with predefined measures in addition to the timely alert. The secondary task of the SA is to furnish an ECP with the job information requested and keep a record of training statistics, including the total number of errors, frequent errors, and the time taken to execute an exercise. To accomplish all these tasks, we reviewed 29 existing autonomous systems such as Coactive Design, ACT-R cognitive architecture, and the intelligent-Coalition Formation (Johnson et al., 2014). Nine essential functional blocks, including interface, system state control, arbitration, information (processing, collection, and storage), goal recognition and mission planning, dynamic task allocation, rules and roles, validation and verification, and training, were identified throughout the detailed review.

As shown in figure 1, the architecture comprises a combination of a natural language processor (NLP) and a speech synthesizer (SS) playing the role of the interface and a database block. The database block is further comprised of an action sequence, control sequence, and user profile databases. The primary operation block contains a data pre-processor, post-processor, scenario calculator, process monitor, and result synthesizer functions. The database block provides information on mission planning, rules, and roles whereas the result synthesizer serves the functionality of goal recognizer, system state controller, and dynamic task allocator.

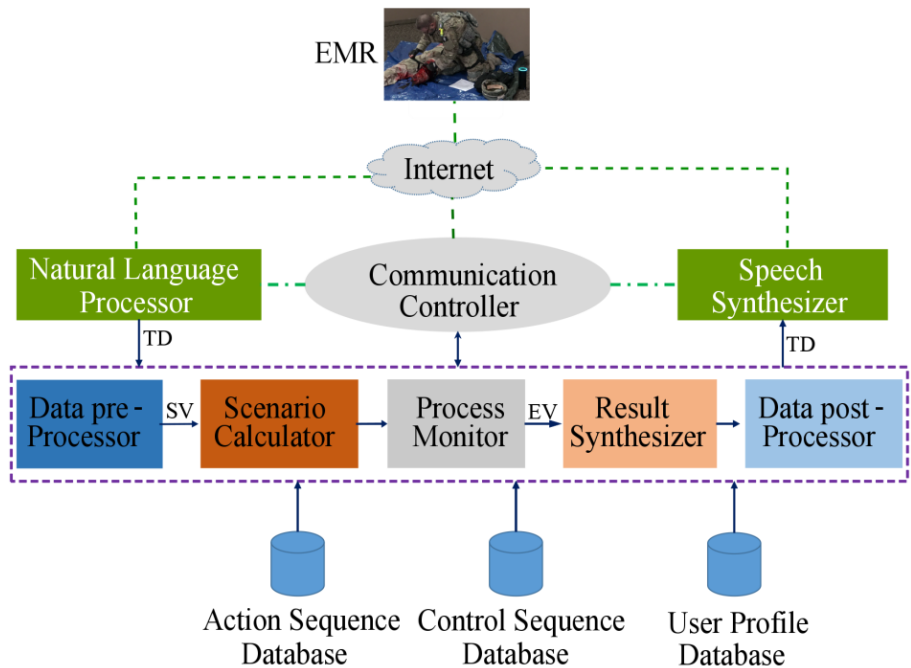

**Fig. 1.** Voice-Based Synthetic Assistant (SA) Architecture

The communication controller manages communication among functional blocks wherein NLP converts speech information into Text Data (TD). The data pre-processor extracts Scenario Variables (SV= process title names) from the TD received. Furthermore, the scenario calculator collects SV and determines the process name to initiate a sequence for the process monitor (PM). The primary function of the PM is to monitor each step of a defined process based on the TD, and SV received. In addition, the process monitor creates an error variable (EV) on the notice of errors in a user step and will be passed to a result synthesizer that successively determines the text. The result synthesizer sends the text to the data post-processor to construct appropriate sentences. Constructed sentences are then passed onto speech synthesizer, which in turn relays this speech to the user through a headphone or speaker.

Table 1 represents a comparative analysis of various technologies and developed VBSA to understand their relevance with one another. The technologies mentioned in the analysis are selected based on their application on the training of various medical experts. For example, the Laerdal robotic manikin is a commercial product for medical training (Jeffries, 2009); Combat Medic virtual training is a gaming software for studying paramedic training in military research labs (Graafland, Schraagen, & Schijven, 2012); and multiple studies of custom pedagogical agents exist in the medical field (Bertrand, Babu, Polgreen, & Segre, 2010; Spector, Merrill, Elen, & Bishop, 2014). Most of the robot manikin technologies mentioned here do not have a voice interface and primarily depend on direct touch and hearing-based interaction (hands-on manikin) to simulate a response without sequential step monitoring. However, with the integration of sensory interaction and nerve system simulation add-ons, the robot manikins can measure trainee implementation with higher accuracy than the rest of them. On the other hand, game-based virtual simulators were considered ineffective by user studies (Barnes et al., 2016; Graafland et al., 2012) for training due to a lack of physical interaction and perception (hands-on practice and natural language communication). On the other hand, virtual pedagogical agents are advanced computer-generated avatars that can also monitor a process. On that note, our VBSA can be termed as a pedagogical agent without a computer avatar, a scaled-down version. Although the developed VBSA only interacts via voice, its architectural framework is developer friendly and exhibits compatibility to integrate with other technologies mentioned above, through APIs. For example, VBSA can be integrated with a robotic manikin, providing manikin sensor data to the data processor in our architecture to recognize task implementation performance through its process monitoring and database management.

Table 1. Comparison of developed VBSA with technologies in practice.

| Features / Medic training technology | Voice Interaction Level | Process Monitoring Capacity | Error Correction Methods | Analytics | Scalability | Integration Capability |
|---|---|---|---|---|---|---|
| **Voice Based Synthetic Assistant (VBSA)** | Natural language based voice interface with complex sentences | Self-reported step by step process monitor | Voice and visual error alert system | Trainee analytics with training history e.g. frequent errors | Yes, and accessible through the internet | Can be integrated with prominent medical systems |
| **Laerdal Robotic Manikins** | No natural language interface | Step by step process is not monitored | Errors alert is well performed though audible sounds and data | Training analytics are provided to trainer | No need for physical hardware | Yes, API available made it possible to integrate this. |
| **Combat Medic Virtual Training** | No Voice | Process monitor through gaming | Error reporting through scoring | Provides trainee analytics for process | Yes | Data not available |
| **Custom Pedagogical Agents in Medical Field** | Natural language interface is available in various levels. | Feature is available, but an emergency medic training is not implemented. | Visual and voice-based error correction methods | Trainee analytics can be obtained. | Yes | API allows them to integrate with most available software systems. |

*2.2 Development*

A cloud platform offers versatility and robustness in addition to profound accessibility in voice-based applications. To date, many cloud service providers provide tools required including NLP and SS features. Amazon Web Services (AWS), Google Cloud, and IBM Watson are few examples to notice. With regard to language features and micro-services available, the developed SA technology is built into AWS. Most of the medical training facilities are equipped with wireless local area network access. This service can be used in most of the training facilities. Still, one of the major problems using cloud systems is a loss of connection or communication interruption. In recent years, the frequency of this problem is reduced significantly. Communication interruptions (loss of internet) will make this service unavailable for training.

Figure 2 shows an SA model utilizing the AWS cloud service to interact with a user. The proposed model uses Alexa Voice Service (AVS) containing NLP, and SS with an off-shelf smart speaker and mobile application interface called Echo. Although AVS can be trained for different accents without any loss in accuracy, a standard American accent was chosen for the experimental purpose. The database was developed and stored in Amazon Dynamo DB, and primary operation blocks (scenario calculator, process monitor, result synthesizer) are developed in AWS Lambda. The user can access the VBSA using the Echo smart speaker or the Amazon Alexa mobile application. Figure 3 shows an actual demonstration of the developed technology where a paramedic is attempting to treat an injured person (manikin with massive hemorrhage on the left leg) with the help of a VBSA design using Amazon Echo (shown with a blue ring near the right bottom corner). The complete demo video can be watched online using the link provided in the picture caption.

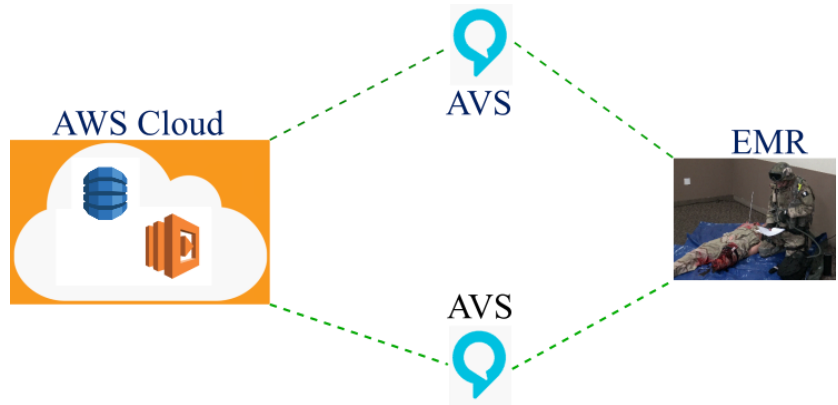

**Fig. 2.** Cloud Based SA Model

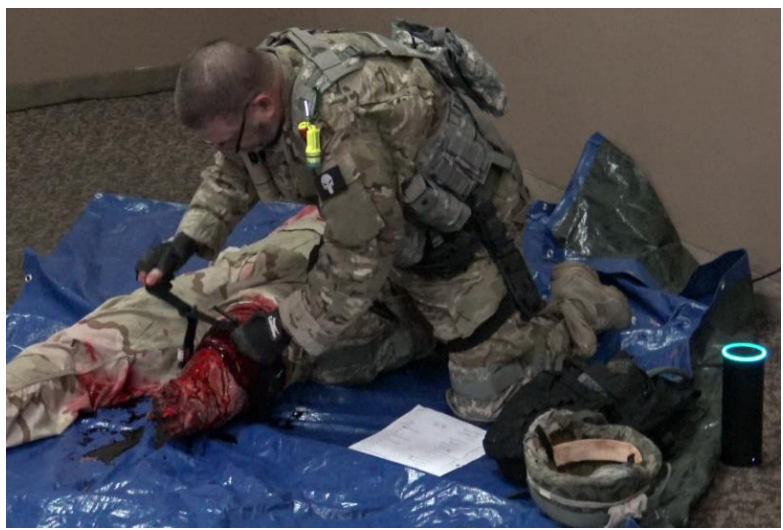

**Fig. 3.** Demo: Paramedic Treating Model Patient with Massive Hemorrhage (https://youtu.be/MNJYfnMta0s)

One of the critical questions that may arise with the use of these technologies is storage of medical information in compliance with the Health Insurance Portability and Accountability Act of 1996 (HIPAA). According to privacy rules of HIPAA Title II, all medical health records of individuals are protected and are to be stored only in a secure location (Terry, 2009). HIPAA also issued guidelines for cloud computing services to comply with HIPAA regulations by implementing electronic-protected health information (ePHI) by a cloud service provider that allows storing electronic health records through encryption and password protection on Amazon servers (Schweitzer, 2012). To ensure HIPAA compliance of our technology, we followed the standard guidelines by Amazon to achieve HIPAA compliance (Amazon Web Services, 2017). Above all, the technology in current usage is aimed at only training and uses fabricated patient health records or information.

*2.3 Working and Usage*

Currently, the developed VBSA works on the MARCHp protocol and can be modified to support any training process in emergency medicine. As part of the process, a trainee uses an interface to interact with VBSA, such as Amazon echo or an android application, and our system supports both interfaces. On the other hand, a trainer uses a web-based dashboard, as shown in figure 4, to monitor the training progress of all the trainees simultaneously, whereas the administrative staff uses a web-based form as an interface to build new training scenarios. The VBSA acknowledges the voice command of the trainee with voice feedback to commence the testing scenario.

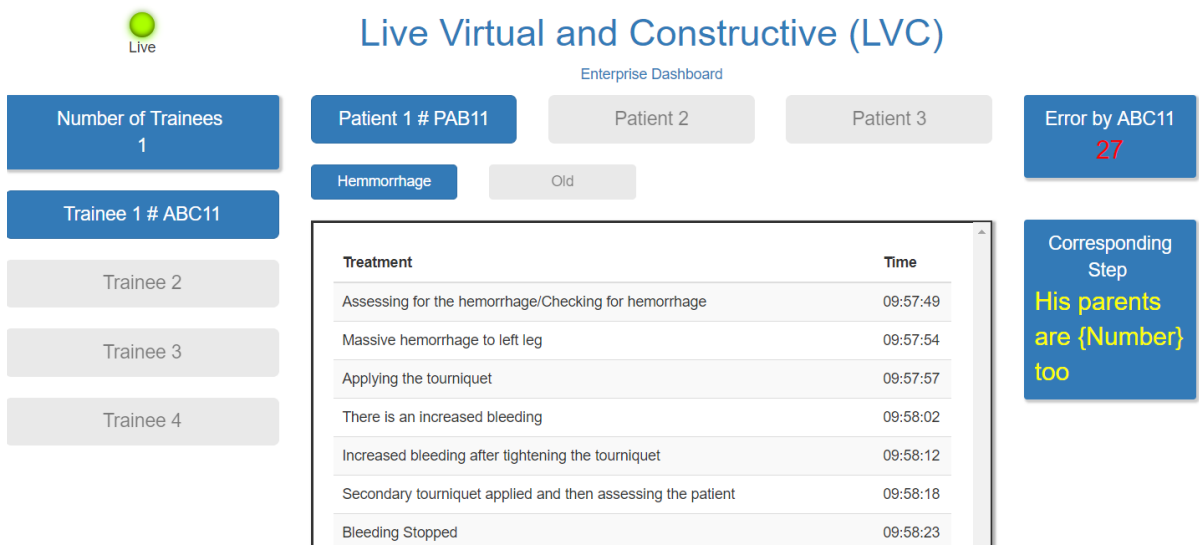

**Fig. 4.** Web-Based Dashboard for Trainer to Monitor Trainees Progress

For example, the trainee initiates a voice command, such as "Alexa, ask assistant," and VBSA acknowledges it with either a voice feedback like "Okay" or else remains silent. In the background of the conversation, the voice command received by the Echo dot is converted into text through NLP. The data pre-processor analyzes the text to extract keywords, such as "assistant" and relays output to the subsequent section. Here, the scenario selector selects a scenario from its database, such as "hemorrhage treatment" based on keywords extracted in the previous section. As a follow-up, the scenario selector assigns variables to point execution steps in hemorrhage treatment and then passes the scenario variables to the process monitor. The process monitor tracks the trainee's task execution in the treatment process and keeps a record of time and errors, such as missing steps. The result synthesizer inspects the error variables related to the steps involved in the hemorrhage treatment and generates a text. Subsequently, the data post-processor constructs text responses that match input from its database and passes the text output to a speech synthesizer. Finally, it is the job of the speech synthesizer to convert the text data into a comprehensive voice response and feedback to the trainee on treatment status. An example conversation between VBSA and a trainee in treating a massive hemorrhage scenario is shown in figure 5. In addition, a decision tree diagram approach is depicted in figure 6 to illustrate a typical VBSA operation.

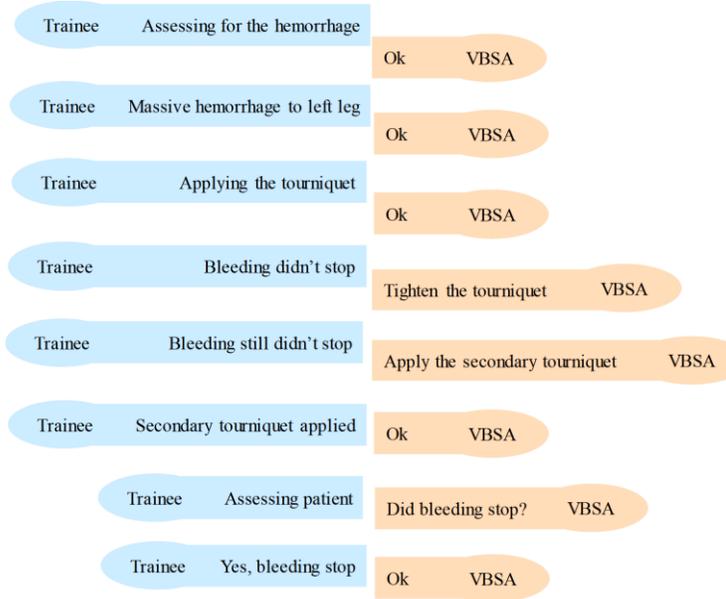

**Fig. 5.** An Example Conversation in Treating Massive Hemorrhage Scenario

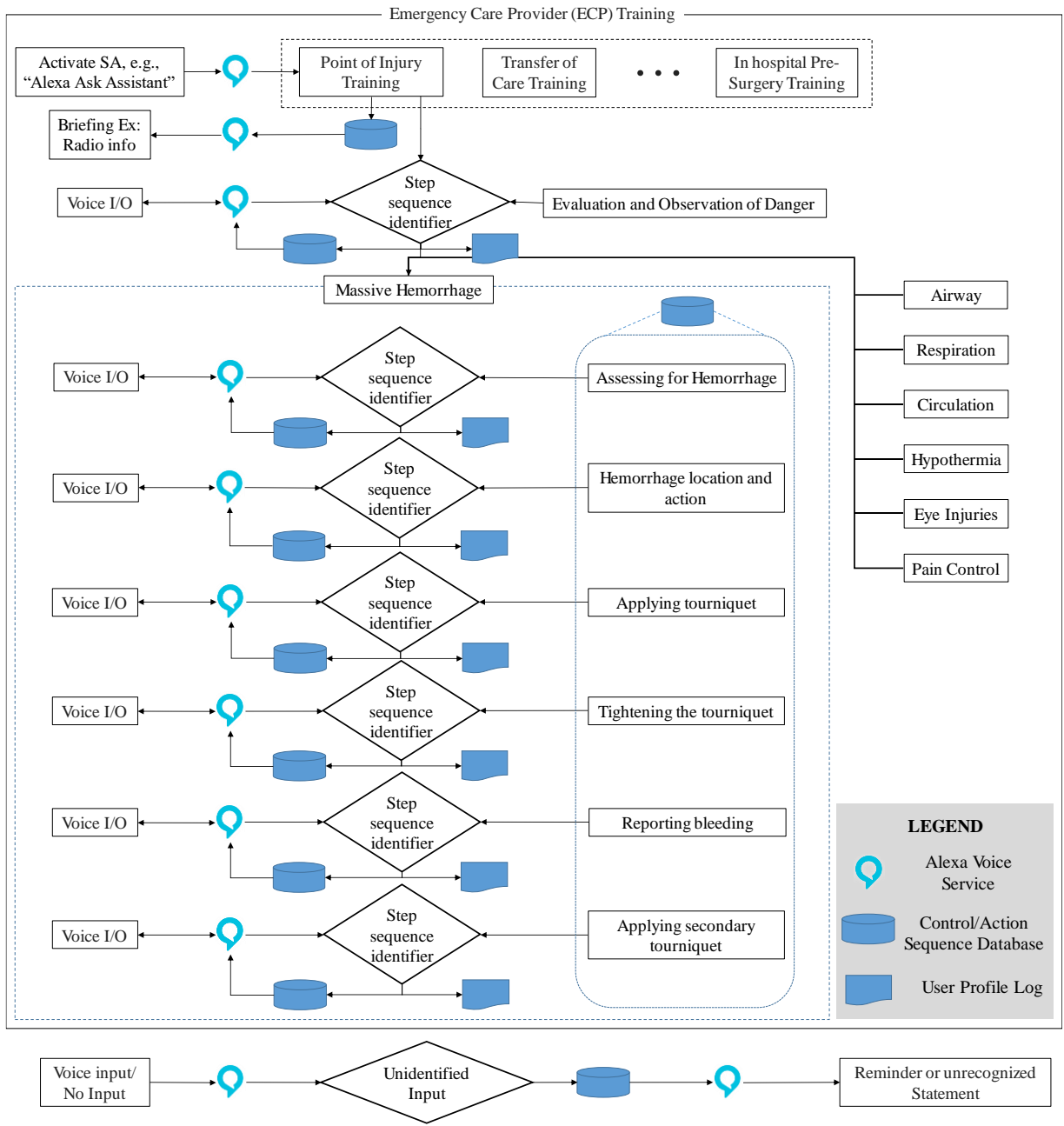

**Fig. 6.** Partial Decision Tree Diagram of VBSA Operation for ECP training with Hemorrhage Scenario

## 3. Method

The research study was piloted through an experimental setup to scrutinize and validate the effectiveness of the developed VBSA model as an interactive training companion to assist an individual in executing critical medical tasks. The discussion of the study approach is divided into six subsequent sections: (1) Testing scenario, (2) Experimental setup, (3) Participants, (4) Procedure, (5) Data collection, and (6) Data analysis. The experiment was conducted on two novice population sets to analyze the performance changes in an individual. The population sets are categorized into two types; a) individuals trained with conventional methods (called a control set) and b) individuals trained with VBSA (called an experiment set). In the due process, experimental performance metric parameters (task execution time, implementation errors, recording errors, and subjective cognitive load for the participant; true positive, true negative, false positive, and false negative interactions for VBSA) were collected to learn the effect of the VBSA in training an individual. Tracking performance and error detection ability of the developed VBSA were assessed to learn the relation between the individual performance and VBSA capabilities.

*3.1 Testing Scenario*

The paper presents a massive hemorrhage treatment scenario selected from the MARCHp protocol of military medic first responder training (Alison Kabaroff, 2013; Kragh Jr et al., 2009). An established scenario for the selected activity is as follows: "A patient was wounded badly on his left leg with heavy bleeding and needed an immediate medical attention by an ECP." The process is to "stop the bleeding by properly applying two tourniquets on the leg within a given period from the time of task commencement." The task was designed to stop bleeding in hemorrhage treatment, and the selected process was divided into 11 to 13 steps of implementation for the participant (acting as an ECP trainee) to follow. The process starts with the paramedic arriving at the scene to assess hemorrhage and ends with a successful application of the second tourniquet to stop bleeding. As a point to note, a simple procedure was chosen for hemorrhage treatment in the experiment. It did not cover an actual process which usually entails a complete list of medical details as well as training material to treat hemorrhage. However, an SA cannot replace a trainer and is only used to enhance a training program.

*3.2 Experimental Setup*

The experimental setup comprises a leg manikin model with a massive hemorrhage and an Amazon Echo working as the VBSA desktop interface, as shown in figure 7(a) that facilitated the setup of monitoring metrics to oversee the experiment arranged in a quiet room. For the experiment set of participants, the instructions for using Amazon Echo was provided in print with a leg manikin model, as shown in figure 7(a). For the control set of participants, a leg manikin model and steps for treatment were provided, as shown in figure 7(b). The leg manikin used in the experiment was not a complete manikin, but was simply developed based on usage. In the pre-experiment briefing, each participant was

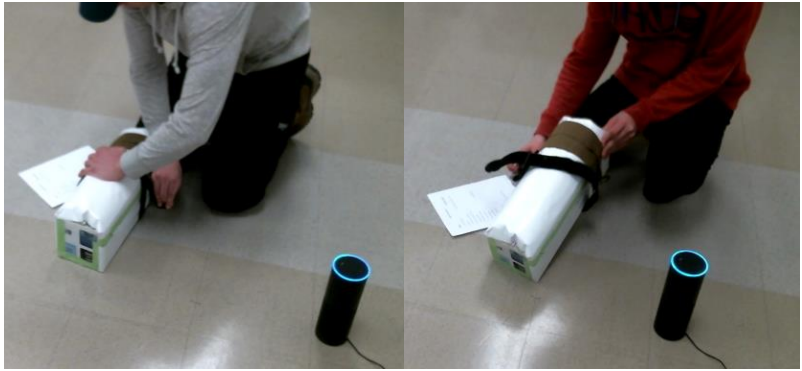

**7(a).** Experiment set participants training with a VBSA

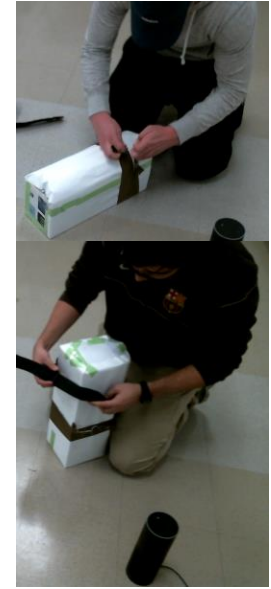

**7(c).** Testing participants

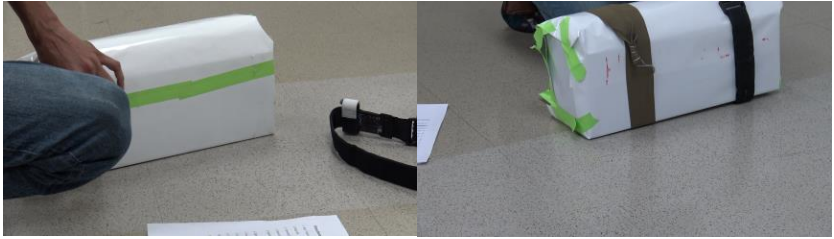

**7(b).** Control set participants training without VBSA

**Fig. 7.** Experimental Setup: Participants Treating Model Patient Leg with Massive Hemorrhage

briefed about the leg manikin setup. The experiment was arranged in a quiet room with a separate provision for a briefing. Time-stamped video was also recorded with participants' identity discretion. In the experiment, defined performance metrics parameters (task execution time, implementation errors, recording errors, and subjective cognitive load for the participant; true positive, true negative, false positive, and false negative interactions for VBSA) were recorded independently by two trained observers and VBSA to avoid any human errors. If recorded values disagreed in any level among the observers or with VBSA, then those values were resolved by revisiting the recording of the time-stamped video recorder.

*3.3 Participants*

A total of thirty-six participants were recruited from the University of Toledo and a local high school. The median of these thirty-six participants' ages was 22 years (a minimum of 19 years and a maximum of 36 years); seventeen participants were high school graduates, sixteen participants hold bachelor degrees, and two hold master degrees. Among the thirty-six participants in the experiment, fourteen were female. The selection of volunteers was based on English accent to minimize technology bias in the performance of the individual with the VBSA. The test involved reading four different sentences. If the designed VBSA recognized all four at the first attempt, then it was a 100% score. The participants who scored 100% in English accent proficiency were taken into the experiment group, and the rest of the participants were assigned to the control group. This English accent proficiency test is to minimized VBSA bias due to NLP accent limitations. We assigned participants to two population sets, the experiment set containing 19 and the control set containing 17 participants. All the participants were novices and did not know MARCHP protocol or hemorrhage treatment in general before participating in this experiment. This gives an opportunity to study the learning curve of a participant and the effect of VBSA on participant's training. Moreover, to avoid any pre-knowledge biases, all the participants were provided with the same basic information related to hemorrhage treatment before the experiment.

*3.4 Procedure*

It is imperative to develop an experimental procedure with predefined guidelines to ensure the successful implementation of a live scenario. The devised experiment was based on Anderson's 2002 work on studying and understanding human cognition and was grounded in Newell's Timescale of Human Activity (Anderson, 2002, 2009; Newell, 1994). The total duration of the experiment was 10 - 14 minutes in varying intervals, as established by Newell's, in cognitive performance testing. Moreover, extreme care had been taken to ensure that the participant was a novice and had no prior knowledge of the testing scenario, even forbidding any contact with a fellow participant in the training and testing session. Initially, the control and experimental population sets were briefed individually with the task description narration, followed by a training and testing section for each participant.

- Participants from the control population set were trained to treat a massive hemorrhage, as described in section 3.1 with conventional methods (an instruction sheet beside the trainee in figure 4(b)), as practiced in conventional training to acclimatize themselves with the task.
- Participants from the experimental population set were trained with the help of VBSA (Amazon Echo acting as VBSA and an instruction sheet shown in figure 4(a)) for the same massive hemorrhage treatment.
- After the brief training session, both control and experiment participants performed the scenario on a mannequin model leg without any instructions from the examiner or the VBSA, as shown in figure 4(c) recorded as testing.

Each training and testing session was separated by a considerable time lag to reduce load and information bias among participants. The experimental procedure was iterated thrice, and therefore, supplied three sets of training and testing data. The task participants trained was to fasten the first tourniquet tightly around the point of injury and check the bleeding. Subsequently, the secondary tourniquet was employed to stop the massive hemorrhage completely and bring the patient to a stable condition. The action performed by the individual was recorded upon verbal and visual confirmation in each step. The critical period set to save the patient was 65 to 95 seconds, and the ideal time was 40 seconds, as established by a subject matter expert. During the testing phase, participants from both populations received no assistance by any means and were expected to perform alone. The training and testing periods of both population sets were video recorded for any further evaluation, as per standard data collection guidelines with the test results garnered in a secure system.

*3.5 Data Collection*

The metrics collected from a participant included time of completion, implementation errors, reported errors, and a corresponding number of the error step. Consequently, this enabled mathematical calculation of a performance score, known as the P-score of each participant in the experiment. Indeed, several procedures were implemented to collect each data point with the least number of errors related to data collection. Two recorders collected the data simultaneously; also, a timer was implemented in terms of timestamps to record the data point at each testing and training period. The video recordings were also gathered for subjective analysis of individual and VBSA performance. However, the data collected from an experimental population set in training also included a number of interventions by the VBSA, the rate of response, and the tracking performance of the VBSA to the aforementioned data. The SA used in the experiment cannot distinguish error types corresponding to tourniquet application and only render information about time of execution and number of missing steps committed by the trainee in a scenario. Furthermore, the VBSA asks the trainee to confirm bleeding at each step to assess the implementation sequence and decide execution efficacy. However, the experiment is monitored additionally by two human observers in real time recording and reporting errors and execution errors. As a point to note, field errors were reduced by sticking to a strict code of communication and proper lexicon training.

*3.6 Data Analysis*

Through this study, we wanted to determine: (a) if training performance is improved due to VBSA, (b) the effect of VBSA on the speed of task execution, and (c) how the VBSA affects human performance and accuracy of treatment (e.g., reduction of errors made by VBSA in training). From the recorded metrics, each participant's Pscore was calculated based on Equation 1. The score was assessed by subtracting penalties from a constant number, where each penalty was calculated based on the importance of the task suggested by subject matter experts. The formula is shown below in Table 2 and penalty for time providing the possible penalty values we can get in this experiment. This formula established in this paper helps in statistical analysis that requires generating a unique individual score for every iteration.

$$Pscore = 10 - (Penalty\ for\ time + \sum_{i=1}^{13}(Penalty\ for\ implementation\ error + Penalty\ for\ recording\ error)) \qquad (1)$$

$$Penalty\ for\ time \begin{cases} if\ task\ time\ (t) < 50, penalty\ =\ 0 \\ if\ 50 < t < 95, penalty\ =\ (t - 40) * 0.02 \\ if\ t > 95, penalty\ =\ 10\ (task\ failure) \end{cases}$$

Table 2. Step implementation and reporting error penalty table

| Step Nos ($i$) | Step description | Penalty for implementation error | Penalty for recording error |
|---|---|---|---|
| 1, 2, 7, 9, 11, 12, 13 | Checking, assessing, re-assessing, reporting | 1 | 0.5 |
| 3, 6 | Applying two tourniquets | 2 | 2 |
| 4, 5, 8, 10 | Tighten tourniquets, check for bleeding | 1.5 | 1 |

The rationale for different steps having different weights are: (1) each steps error leads to a different level of severity, for example, improper application of a tourniquet is more severe than re-assessment step, and (2) implementation error has more weight than reporting an error in most of the cases. A 10-second adjustment was made as per subject matter expert suggestions in the formulation to get the baseline performance score of novice participants for this experimentation. An example Pscore calculation in training/testing for experimental/controlled is if participant executed all 13 steps in 65 seconds and made errors in the implementation of step 8, then $Pscore = 10 - ((65 - 40) \times 0.02 + 1.5) = 8$.

Data for participants considered for Pscore calculation only if he or she legitimately performed all experiment steps. Pscore and the total number of errors in both experimental and controlled groups, for both testing and training cases, are statistically analyzed to see if VBSA improves the performance. Welch's t-test was used to test the significance of differences in Pscores of the control group and the Experiment group (Ruxton, 2006). The Wilcoxon Signed-Rank test was used to test the significance of the improvement in performance scores from test 1 to test 3 and from train 1 to train 3 in the same group (Gibbons & Chakraborti, 2011); to break the ties, we used the approach defined by (Pratt, 1959).

In a subjective analysis of data, the experimental study also considered the participants' feedback over the accessibility of SA in terms of natural language and participants' degree of comfort with SA. The examiner observes language exchange and the timing between participant and SA during both training and testing. Even though assessment included positive and negative scale (Watson, Clark, & Tellegen, 1988), limitations of this analysis were considered (Naismith, Cheung, Ringsted, & Cavalcanti, 2015).

## 4. Results and Discussion

### 4.1 Developed VBSA Results

The primary goal of VBSA is real-time monitoring of ECP trainee with error detection and recommendation of steps to resolve errors. Besides that, SA is also designed to record task execution time and errors. figure 8 shows implementation and VBSA operation results, and the video link provided shows the successful operation of VBSA using experiment test case. At the same time, figure 8(a) is a screenshot of Amazon Dynamo DB database containing MARCHP process steps, figure 8(b) is a snippet of code developed in Amazon Lambda that is responsible for primary blocks operation. figure 8(c) is a screenshot of the AVS NLP interface designed for MCR training for the MARCHP process.

**Fig. 8(a)** Treatment process steps developed in DynamoDB Database

**Fig. 8(b)** Sample VBSA Processing code          **Fig 8(c)** Interface design code for NLP

**Fig. 8.** VBSA Development and Operational Results Screenshots

The interaction effort of VBSA can be represented by observing true positive (TP), true negative (TN), false positive (FP), and false negative (FN) metrics. Every interaction or user speech command to VBSA can be categorized in one of those four classes as a stationary value with no probability involved. If a scripted speech command is performed accurately and VBSA executes it accurately, then it is TP. If a scripted speech command is performed accurately, but VBSA rejects it, then it is FN. If a scripted speech command is performed inaccurately and VBSA rejects it, then it is TN. Finally, if a scripted speech command is performed inaccurately, but VBSA accepts any step based on it, then it is FP. We had 19 participants who belonged to the experimental set that used VBSA for training. Each participant needed to execute 13 steps in each of the three training sessions. Figure 9 presents the interaction metrics of each of these three sessions represented as Train 1, 2 and 3. The number of total 247 TP interactions for each session resulted from the multiplication of total number of participants (19) and the required number of steps to be executed for each individual session (13), i.e., 13 * 19 = 247. Here, TP and TN represent the correct operation of VBSA interface whereas FN and FP represent the error operation of the interface. As we can observe from the graph below, the value of FP is very low and suggests that it is highly unlikely that the VBSA will accept any unwanted command, making the developed system reliable.

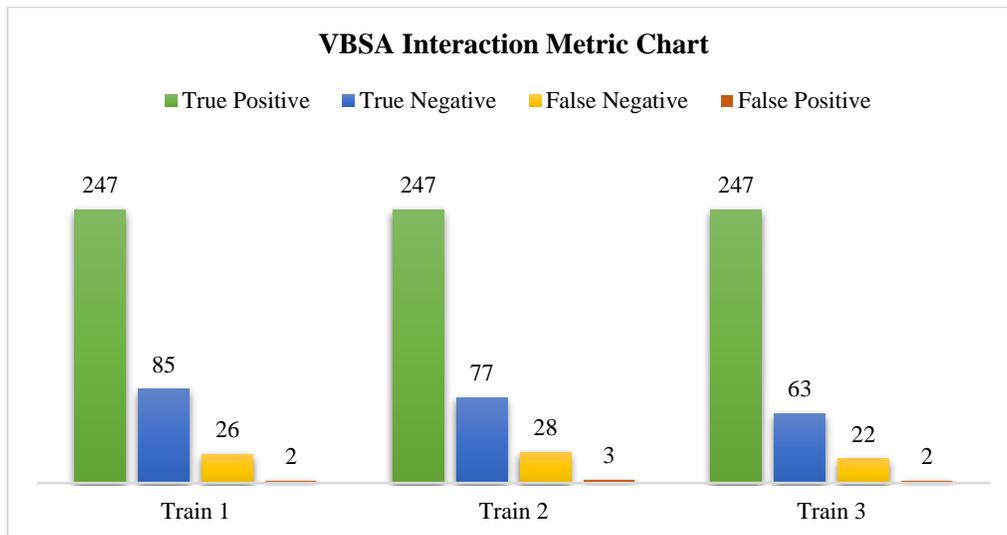

**Fig. 9.** VBSA Interface Metric Chart

*4.2 User Studies Statistical results*

As described in the method section, user studies were performed on developed VBSA to study the effect of it on the performance of medical trainees. Figure 10 represents the performance score (Pscore) for both the control group and the experiment group calculated using equation 1 for all training and testing iterations. Welch's t-test found statistically significant improvement in the Pscore of the experiment group in test 3 compared to the control group in the same test *(z = -2.1538, df = 29.757, p = 0.01974)*. As seen in figure 10, as the training iterations increase, the mean of the Pscore also increases for both the groups.

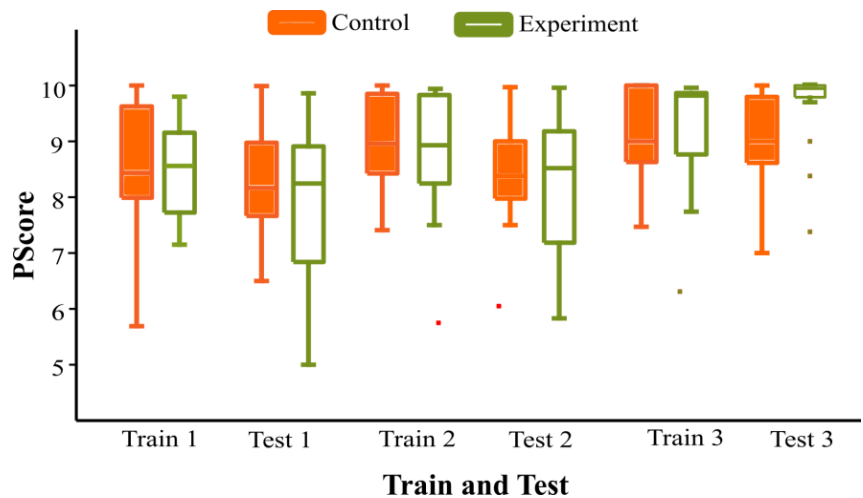

**Fig. 10.** Pscore by iteration of train and test with population groups

Welch's t-test between the control group and the experiment group on test 2 data determined that the Pscore difference is not statistically significant *(p = 0.6395 tested for p < 0.05)*. It was also evident that between the two groups, the Pscore difference was not statistically significant *(p = 0.2295)* for test 1.

The two-tailed Wilcoxon sign ranked the differences between tests 1 and 2 *(z= -0.6905, p = 0.4902 for p < 0.05)*, and between tests 2 & 3 *(z= -1.8743, p = 0.06148 for p < 0.05)* of the control group showed no statistically significant difference in the Pscore. Nevertheless, statistically significant growth existed from test 1 to test 3 *(z = -2.1315, p = 0.03318 for p < 0.05)*. The same test applied on the experimental group between tests 1 and 2 *(z = -2.1976, p = 0.0278 for p < 0.05)* and tests 2 and 3 *(z = -3.5162, p = 0.00044 for p < 0.05)* indicated a statistically significant difference in Pscores, as well as, a statistically significant growth in Pscore from tests 1 to 3 *(z = -3.5162, p =0.00044 for p < 0.05)*.

In conclusion, figure 10 illustrates that even after multiple training sessions, the increase in performance (Pscore) was statistically insignificant for the control group, whereas the Pscore of the experimental group has significantly improved from test 1 to test 2 and from test 2 to test 3, and improved significantly compared to control group. Based on the above test results, the hypothesis of the individual performance enhancement in training with VBSA when compared to conventional training was mostly validated. Furthermore, we noticed that the recursive training turned out to be more effective with significant growth in the successive tests 1, 2, and 3 when coupled with the VBSA than in conventional training. These results may change with large samples and more patient treatment test cases.

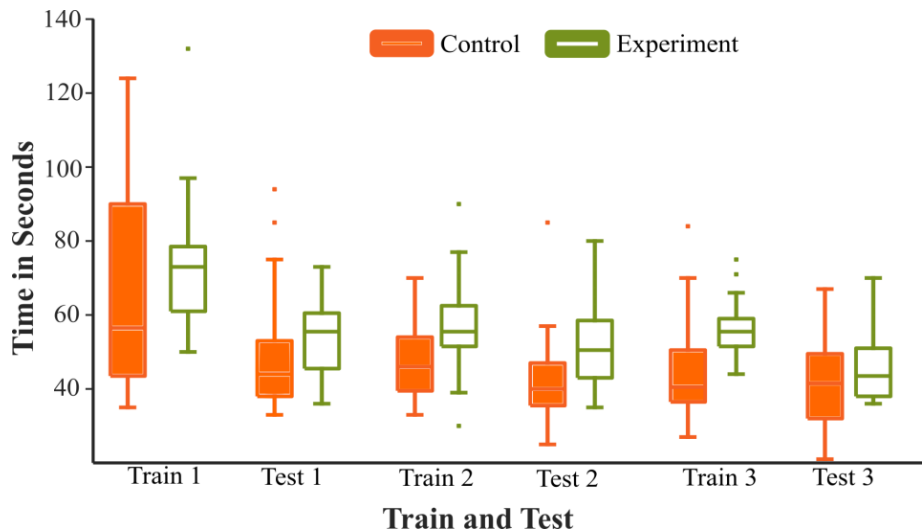

**Fig. 11.** Task time by iteration of train and test with population groups

The total time to complete a given task for the control group, and the VBSA assisted group is presented in figure 11. Welch's t-test showed that there was no statistically significant loss in execution time between the control group and experiment group in any testing scenarios, which showed that the effect of cognitive load on individuals due to the VBSA was minimal. As shown in figure 11 and from the analysis done using two-tailed Wilcoxon sign ranked test *(experimental = (z = -2.6694, p = 0.00758 for p < 0.05), control = (z = -2.8682. The p = 0.0041 for p < 0.05))* validates our hypothesis that an increase in training iterations reduced the time taken in the testing section by both groups, but there is no statistically significant difference between those two. Based on the above data, it is observed that the training time of the experimental group is longer than the time taken by the control group. The discrepancy in training time can be attributed to limited vocabulary interface.

Based on the above experiment results and analysis of the time taken to complete a task, the use of VBSA seems to have no adverse effects on the individual's task execution speed, cognitive load, and stress factor. These results may change with large samples and more patient treatment test cases. The overall distribution of mistakes (All) and the error distribution of each group is summarized in Table 3. As represented in Table 3, our tests and observations show a difference between implementation errors (IMP) and reporting errors (REP) in both the control set and the experimental set. Results show that reporting errors are significantly reduced in the experimental set, which holds well for implementation errors also compared to the control group.

**Table 3.** The two-tailed Wilcoxon sign ranked test result for all PScores in comparison with total number of errors made by individuals in each attempt

|  |  | Train 1 | Test 1 | Train 2 | Test 2 | Train 3 | Test 3 |
|---|---|---|---|---|---|---|---|
| *p* | | <0.05 | <0.05 | <0.05 | <0.05 | <0.05 | <0.05 |
| **Control Errors** | IMP | 15 | 16 | 8 | 14 | 7 | 9 |
|  | REP | 23 | 27 | 13 | 22 | 11 | 11 |
| **Experiment Errors** | IMP | 9 | 15 | 10 | 15 | 9 | 2 |
|  | REP | 16 | 28 | 10 | 16 | 2 | 5 |

Based on Table 3 results and its analysis, the final question posed about accuracy improvement through VBSA training was mostly satisfied by reducing reporting and implementation errors. Results indicate that reporting errors were reduced significantly in each training session in the experiment set compared to the control group. Implementation errors were only reduced from the first set to the last set of training in the experiment group compared to the control group. Comparing the control set training with that of the experiment set using figure 11 and Table 3, we can conclude that implementing a VBSA increases the total training time, which may be related to using scripted language in VBSA, which need further investigation. If further studies with a large sample and more test cases provided any statistically significant effects of scripted language in training ECP that effects field performance (testing), which becomes a major setback to VBSA that needs to be addressed.

5.  **Conclusion**

It is well known that a large number of deaths result from medical errors. It is therefore obvious that during emergencies, emergency care providers (ECPs) are prone to making errors while the accuracy and response time of their actions are critical factors in the treatment of patients. Any improvement in these factors would significantly increase the chances of survival of a patient. On that note, a primary focus of ECP training should be reducing the response time as well as enhancing the accuracy of ECP responses. Since conventional methods may not be sufficient in achieving this goal, we proposed and developed a VBSA and evaluated its effect on ECP training. Lessons learned in developing the proposed VBSA and in testing are as follows: limitations of VBSA rendered by scripted vocabulary had an observable effect on an ECP's interaction with the SA in training that needs further investigation. The possibility of implementing a VBSA in real training is viable only if VBSA is subjected to (1) rapid development for real-time application, (2) verification and validation of VBSA, and (3) sustainability of deployment. To conclude, use of an SA for medical training including several aspects of the treatment process is expected to result in efficient and effective training. As the VBSA is capable of handling multiple trainees, it is expected to minimize training costs and enhance utilization of human resources available for training without compromising the training quality. The SAs are affordable, upgradable, and highly compatible with other existing medical emergency training programs as well as application areas ranging from medicine to industry and education. It can be concluded from current experiment result that the use of a VBSA in ECP training is a reasonable choice to improve performance and cut down associated training costs. In its current form, the deployment of VBSA is not recommended in training due to its limitation, such as lack of large sample studies on effects of VBSA on trainees. Rigorous development and more testing are required to address these and other limitations that are yet to be discovered.

**Acknowledgements**

The University of Toledo and Round 1 Award from the Ohio Federal Research Jobs Commission (OFMJC) through Ohio Federal Research Network (OFRN) fund this research project; authors also appreciate support of EECS department at the University of Toledo.